\documentclass[11pt]{amsart}
\def\isdraft{0}

\usepackage{txfonts}
\usepackage{amsaddr} 
\usepackage{mathtools}
\usepackage{amsmath,amssymb,amsfonts,dsfont}
\usepackage{prettyref} 
\usepackage[utf8]{inputenc}
\usepackage{enumitem}
\usepackage[hyphens]{url}
\usepackage{tikz-cd}
\usepackage{tikz}
\usetikzlibrary{positioning}
\usetikzlibrary{arrows}
\usepackage{bussproofs}
\usepackage[mathscr]{euscript}
\usepackage{hyperref} 
\usepackage{amsthm}
\usepackage{theapa}
\usepackage{a4wide}
\usepackage[color=black,textcolor=white\if\isdraft0,disable\fi]{todonotes}
\usepackage{stackengine}
\usepackage{stmaryrd}
\usepackage{graphicx}
\usepackage{wrapfig}
\usepackage{float}
\graphicspath{{fig/}}

\theoremstyle{definition} 

\newrefformat{cha}{Chapter \ref{#1}}
\newrefformat{sec}{Section \ref{#1}}
\newrefformat{tab}{Table \ref{#1}}
\newrefformat{fig}{Figure \ref{#1}}
\newrefformat{equ}{(\ref{#1})}
\newrefformat{app}{Appendix \ref{#1}}
\newrefformat{thm}{Theorem \ref{#1}}
\newrefformat{cor}{Corollary \ref{#1}}
\newrefformat{prop}{Proposition \ref{#1}}
\newrefformat{lem}{Lemma \ref{#1}}
\newrefformat{fact}{Fact \ref{#1}}
\newrefformat{obs}{Observation \ref{#1}}
\newrefformat{note}{Note \ref{#1}}
\newrefformat{idea}{Idea \ref{#1}}
\newrefformat{trivia}{Trivia \ref{#1}}
\newrefformat{def}{Definition \ref{#1}}
\newrefformat{not}{Notation \ref{#1}}
\newrefformat{con}{Convention \ref{#1}}
\newrefformat{rem}{Remark \ref{#1}}
\newrefformat{exa}{Example \ref{#1}}
\newrefformat{problem}{Problem \ref{#1}}
\newrefformat{claim}{Claim \ref{#1}}
\newrefformat{conjecture}{Conjecture \ref{#1}}
\newrefformat{exe}{Exercise \ref{#1}}
\newrefformat{alg}{Algorithm \ref{#1}}
\newrefformat{err}{Error \ref{#1}}
\newrefformat{que}{Question \ref{#1}}
\newrefformat{ite}{Item \ref{#1}}
\newrefformat{Q}{Q. \ref{#1}}
\newrefformat{warning}{Warning \ref{#1}}
\newrefformat{pseudocode}{Pseudocode \ref{#1}}

\title{Algebraic characterizations of least model and uniform equivalence of propositional Krom logic programs}
\author{
	Christian Anti\'c
}
\address{
	christian.antic@icloud.com\\
	Vienna University of Technology\\
	Vienna, Austria
}

\begin{document}
\begin{abstract} 
	This research note provides algebraic characterizations of the least model, subsumption, and uniform equivalence of propositional Krom logic programs.
\end{abstract}
\maketitle

\section{Introduction}


Propositional Krom\footnote{See \citeA{Krom67}.} logic programs consist of facts and proper rules of the simple form $a\leftarrow b$. Although the class of Krom programs is very restricted, least model and uniform equivalence turn out to be still non-trivial in that setting. This is due to the fact that we can identify every such program with a graph where each rule $a\leftarrow b$ corresponds to an edge from $b$ to $a$, which shows that propositional Krom logic programs are, despite their simplicity, non-trivial objects.

Notions of logic program equivalence are important for understanding the computational behavior of programs, for example in program optimization.

Least model equivalence is the most basic form of equivalence studied in the logic programming literature. Formally, two programs $P$ and $R$ are least model equivalent iff their least models---that is, their logical consequences derivable from facts bottom-up---coincide.

Uniform equivalence is a key notion of equivalence \cite{Maher88,Sagiv88,Eiter03} where two programs $P$ and $R$ are uniformly equivalent iff $P\cup I$ is least model equivalent to $R\cup I$, for every interpretation $I$. It has found applications to program optimization.

In this paper, we study least model and uniform equivalence of propositional Krom logic programs from the algebraic point of view in terms of the sequential composition operation recently introduced by the author in \citeA{Antic21-1} \cite<and see>{Antic21-2}.


In a broader sense, this paper is a further step towards an algebraic theory of logic programming.

\section{Propositional Krom logic programs}

This section recalls propostional Krom logic programs by mainly following the lines of \citeA{Antic21-1}. 

Let $A$ be a finite non-empty alphabet of propositional atoms. A ({\em propositional Krom logic}) {\em program} over $A$ is a finite set of {\em rules} of two forms, elements of $A$ called {\em facts}, and {\em proper rules} of the form $a\leftarrow b$ for some atoms $a,b\in A$. We denote the facts and proper rules of a program $K$ by $f(K)$ and $p(K)$, respectively.

An {\em interpretation} is any subset of $A$. We define the {\em entailment relation}, for every interpretation $I$, inductively as follows: (i) for an atom $a$, $I\models a$ iff $a\in I$; (ii) for a proper rule, $I\models a\leftarrow b$ iff $I\models b$ implies $I\models a$; and, finally, (iii) for a propositional Krom program $K$, $I\models K$ iff $I\models r$ holds for each rule $r\in K$. In case $I\models K$, we call $I$ a {\em model} of $K$. The set of all models of $K$ has a least element with respect to set inclusion called the {\em least model} of $K$ and denoted by $LM(K)$. We say that $K$ and $L$ are {\em least model equivalent} iff $LM(K)=LM(L)$.

We define the ({\em sequential}) {\em composition} \cite{Antic21-1} of $K$ and $L$ by
\begin{align*} 
	K\circ L:=f(K)\cup\{a\mid a\leftarrow b\in K\text{ and }b\in L\}\cup\{a\leftarrow c\mid a\leftarrow b\in K\text{ and }b\leftarrow c\in L\}.
\end{align*} We will write $KL$ in case the composition operation is understood.

We can compute the facts in $K$ via
\begin{align}\label{equ:f(K)}
	f(K)=K\emptyset.
\end{align}

Define the {\em Kleene star} and {\em plus} of $K$ by
\begin{align*} 
	K^\ast:=\bigcup_{n\geq 0} K^n \quad\text{and}\quad K^+:=K^\ast K,
\end{align*} and the {\em omega} of $K$ by
\begin{align*} 
	K^\omega:=f(K^+)\stackrel{\prettyref{equ:f(K)}}=K^+\emptyset.
\end{align*} We have an algebraic characterization of the least model semantics of $K$ given by
\begin{align*} 
	LM(K)=K^\omega.
\end{align*}
 
For every interpretation $I$, we have
\begin{align}\label{equ:IK=I}
	IK=I.
\end{align} Moreover, we have \cite[Theorem 12]{Antic21-1}
\begin{align}
	\label{equ:K_cup_L_M} (K\cup L)M&=KM\cup LM\\
	\label{equ:M_K_cup_L} M(K\cup L)&=MK\cup ML
\end{align} for all propositional Krom programs $K,L,M$.

\section{Least model equivalence}

Recall that two programs $K,L$ are equivalent with respect to the least model semantics iff $K^\omega=L^\omega$. We therefore wish to compute $K^\omega$. We have
\begin{align*} 
	K^\omega=f(K^+)=K^+\emptyset=\left(\bigcup_{n\geq 1}K^n\right)\emptyset=\bigcup_{n\geq 1}(K^n\emptyset)=\bigcup_{n\geq 1}f(K^n).
\end{align*} Let us compute $K^n$. We have
\begin{align*} 
	K^2
		&=(f(K)\cup p(K))(f(K)\cup p(K))\\
		&=f(K)^2\cup p(K)f(K)\cup f(K)p(K)\cup p(K)^2\\
		& \stackrel{\prettyref{equ:IK=I}}=f(K)\cup p(K)f(K)\cup p(K),
\end{align*} and
\begin{align*} 
	K^3=K^2K=f(K)\cup p(K)f(K)\cup p(K)^2K=f(K)\cup p(K)f(K)\cup p(K)^2f(K)\cup p(K)^3,
\end{align*} and
\begin{align*} 
	K^4
		&=K^3K\\
		&=f(K)\cup p(K)f(K)\cup p(K)^2f(K)\cup p(K)^3K\\
		&=f(K)\cup p(K)f(K)\cup p(K)^2f(K)\cup p(K)^3f(K)\cup p(K)^4.
\end{align*} A simple proof by induction shows the general formula for arbitrary $n\geq 1$ given by
\begin{align*} 
	K^n=f(K)\cup p(K)^n\cup \bigcup_{i=1}^{n-1}\left(p(K)^if(K)\right).
\end{align*} This implies
\begin{align*} 
	K^\omega
		&=f(K^+)\\
		&=f\left(\bigcup_{n\geq 1}K^n\right)\\
		&=\bigcup_{n\geq 1}f\left(f(K)\cup p(K)^n\cup \bigcup_{i=1}^{n-1} p(K)^i f(K)\right)\\
		&=\bigcup_{n\geq 1}\left(f(f(K))\cup f(p(K)^n)\cup\bigcup_{i=1}^{n-1}f(p(K)^i f(K))\right)\\
		&=f(K)\cup \bigcup_{n\geq 1}\bigcup_{i=1}^{n-1}p(K)^i f(K)\\
		&=\bigcup_{n\geq 1}\bigcup_{i=0}^{n-1}p(K)^i f(K),
\end{align*} where the fifth identity follows from $f(f(K))=f(K)$ and $f(p(K)^n)=\emptyset$, and the last identity follows from $p(K)^0=1$.

Hence, we obtain an algebraic characterization of equivalence with respect to the least model semantics given by
\begin{align*} 
	K\equiv L \quad\Leftrightarrow\quad \bigcup_{n\geq 1}\bigcup_{i=0}^{n-1}p(K)^i f(K)=\bigcup_{n\geq 1}\bigcup_{i=0}^{n-1}p(L)^i f(L).
\end{align*}

\section{Subsumption equivalence}

Recall that two Krom programs $K,L$ are called {\em subsumption equivalent} \cite<cf.>{Maher88}---in symbols, $K\equiv_{ss} L$---iff $KI=LI$ holds for each interpretation $I$.\footnote{Subsumption equivalence is usually defined in terms of the van Emden Kowalski operator...}

Our first observation is that in case $K$ and $L$ are subsumption equivalent, we must have
\begin{align*} 
	K\emptyset=L\emptyset
\end{align*} which by \prettyref{equ:f(K)} is equivalent to
\begin{align*} 
	f(K)=f(L).
\end{align*} Now let $a$ be an atom in the head of $K$ which is not a fact of $K$, that is,
\begin{align*} 
	a\in h(K) \quad\text{and}\quad a\not\in f(K).
\end{align*} Then there must be a rule $a\leftarrow b\in K$, for some $b\in A$. Let $I:=\{b\}$. Since
\begin{align*} 
	KI=LI \quad\text{and}\quad a\in KI \quad\text{implies}\quad a\in LI,
\end{align*} we must have $a\leftarrow b\in L$. This immediately implies $K\subseteq L$. An analogous argument shows $L\subseteq K$, which leads to the trivial characterization of subsumption equivalence given by
\begin{align*} 
	K\equiv_{ss} L \quad\text{iff}\quad K=L.
\end{align*}


\section{Uniform equivalence}

Recall that two Krom programs $K,L$ are called {\em uniformly equivalent} \cite{Maher88,Sagiv88,Eiter03}---in symbols, $K\equiv_u L$---iff $K\cup I\equiv L\cup I$ holds for every interpretation $I$. The condition $K\cup I\equiv L\cup I$ can be rephrased in terms of the omega operation as $(K\cup I)^\omega=(L\cup I)^\omega$. This means that in order to understand uniform equivalence, we need to understand $(K\cup I)^\omega$. 

First, we have
\begin{align*} 
	(K\cup I)^2=(K\cup I)(K\cup I)\stackrel{\prettyref{equ:K_cup_L_M},\prettyref{equ:M_K_cup_L}}=K^2\cup KI\cup IK\cup I^2\stackrel{\prettyref{equ:IK=I}}=K^2\cup KI\cup I
\end{align*} and
\begin{align*} 
	(K\cup I)^3=(K^2\cup KI\cup I)(K\cup I)=K^3\cup KIK\cup IK\cup K^2I\cup KI^2\cup I^2 \stackrel{\prettyref{equ:IK=I}}=K^3\cup K^2I\cup KI\cup I.
\end{align*} A straightforward induction proof shows that the general pattern for arbitrary $n\geq 1$ is
\begin{align*} 
	(K\cup I)^n=K^n\cup K^{n-1}I\cup\ldots \cup KI\cup I.
\end{align*} Hence, we have
\begin{align*} 
	(K\cup I)^+=\bigcup_{n\geq 1}(K\cup I)^n=\bigcup_{n\geq 1}(K^n\cup K^{n-1}I\cup\ldots \cup KI\cup I)=K^+\cup K^\ast I,
\end{align*} and therefore
\begin{align*} 
	(K\cup I)^\omega=f((K\cup I)^+)=f(K^+\cup K^\ast I)=f(K^+)\cup f(K^\ast I)=f(K^+)\cup K^\ast I=K^\omega\cup K^\ast I.
\end{align*} That is, we can compute the least model of $K\cup I$ in terms of the least model of $K$. This implies the characterization of uniform equivalence given by
\begin{align*} 
	K\equiv_u L \quad\Leftrightarrow\quad K^\omega\cup K^\ast I=L^\omega\cup L^\ast I.
\end{align*} This means that in order to understand uniform equivalence, we need to understand the omega and star operators. 

We wish to understand $K^\ast I$. We therefore compute
\begin{align*} 
	K^\ast I
		&=\left(\bigcup_{n\geq 0}K^n\right)I\\
		&=\bigcup_{n\geq 0}(K^nI)\\
		&=\bigcup_{n\geq 0}\left(f(K)I\cup p(K)^nI\cup\bigcup_{i=1}^{n-1}p(K)^if(K)I\right)\\
		&=f(K)\cup\bigcup_{n\geq 0}\left(p(K)^nI\cup\bigcup_{i=1}^{n-1}p(K)^if(K)\right)\\
		&=f(K)\cup\bigcup_{n\geq 0}p(K)^nI\cup\bigcup_{n\geq 0}\bigcup_{i=1}^{n-1}p(K)^if(K)\\
		&=K^\omega\cup\bigcup_{n\geq 0}p(K)^nI\\
		&=K^\omega\cup p(K)^\ast I.
\end{align*} 

In total, we therefore have
\begin{align*} 
	(K\cup I)^\omega=K^\omega\cup K^\ast I=K^\omega\cup p(K)^\ast I.
\end{align*} 

Hence, we have finally arrived at an algebraic characterization of uniform equivalence of propositional Krom logic programs given by
\begin{align*} 
	K\equiv_u L \quad\Leftrightarrow\quad K^\omega\cup p(K)^\ast I=L^\omega\cup p(L)^\ast I,\quad\text{for every interpretation $I$.}
\end{align*}

\section{Conclusion}

This research note provided algebraic characterizations of least model and uniform equivalence of propositional Krom logic programs in terms of the sequential composition of programs recently introduced by \citeA{Antic21-1}.

The major line for future research is to lift the results of this paper from Krom to arbitrary propositional-,  first-order-, and answer set programs containing negation as failure \cite{Clark78}. The former is challenging since the sequential composition of arbitrary propositional programs is in general not associative and does not distribute over union \cite<cf.>{Antic21-1} which means that most of the results of this paper are not directly transferable. The latter is difficult since the sequential composition of answer set programs is rather complicated \cite<cf.>{Antic21-2}.

\if\isdraft0


\section*{Conflict of interest}

The authors declare that they have no conflict of interest.

\section*{Data availability statement}

The manuscript has no data associated.
\fi

\bibliographystyle{theapa}
\bibliography{/Users/christianantic/Bibdesk/Bibliography,/Users/christianantic/Bibdesk/Publications,/Users/christianantic/Bibdesk/Preprints,/Users/christianantic/Bibdesk/Under_construction,/Users/christianantic/Bibdesk/Notes}
\end{document}